\documentclass[12pt]{article}
\pdfoutput=1

\usepackage{amssymb,amsmath,mathrsfs,epstopdf,slashed,color}

\usepackage{ifpdf}
\ifpdf
\usepackage{graphicx}
\usepackage{hyperref}
\usepackage{cite}
\else
\usepackage[dvipdfmx]{graphicx}
\usepackage[dvipdfmx]{hyperref}
\usepackage[numbers]{natbib}
\usepackage{hypernat}
\fi

\allowdisplaybreaks[1]

\makeatletter

\@addtoreset{equation}{section}
\makeatother

\newcommand{\Vol}{\mathcal{V}}

\DeclareMathOperator*{\im}{Im \,}

\title{\bf Probability Comparison}
\author{Yoske Sumitomo, Markus Rummel}

\begin{document}

\begin{titlepage}

\setcounter{page}{0}
  
\begin{flushright}
 \small
 \normalsize
\end{flushright}

\vskip 3cm
\begin{center}

{\Large \bf Probability of Vacuum Stability\\ in Type IIB Multi-K\"ahler Moduli Models}  

\vskip 2cm
  
{\large Markus Rummel${}^{1}$, Yoske Sumitomo${}^{2}$, }
 
 \vskip 0.6cm

${^1}$Rudolph Peierls Centre for Theoretical Physics, University of Oxford,\\ 1 Keble Road, Oxford, OX1 3NP, United Kingdom\\
${}^2$ Institute for Advanced Study and Department of Physics,\\ The Hong Kong University of Science and Technology, Clear Water Bay, Hong Kong\\

 \vskip 0.4cm

Email: \href{mailto:markus.rummel@physics.ox.ac.uk, yoske@ust.hk}{markus.rummel at physics.ox.ac.uk, yoske at ust.hk}

\vskip 1.0cm
  
\abstract{\normalsize
 We study the probability that all eigenvalues of the moduli mass matrix at extremal points are positive in concrete multi-K\"ahler moduli models of type IIB string theory compactifications in the large volume regime.
 Our analysis is motivated by the open question if vacua which are uplifted to de Sitter remain stable.
 We derive a simple analytical condition for the mass matrix to be positive definite, and estimate the corresponding probability in a supersymmetric moduli stabilization model along the lines of KKLT and a non-supersymmetric Large Volume Scenario type of model, given a reasonable range of compactification parameters.
 Under identical conditions, the probability for the supersymmetric model is moderately higher than that of the Large Volume Scenario type model.
}
  
\vspace{1cm}
\begin{flushleft}
 \today
\end{flushleft}
 
\end{center}
\end{titlepage}

\setcounter{page}{1}
\setcounter{footnote}{0}

\tableofcontents

\parskip=5pt

\section{Introduction}

We are now in the era of the accelerated expanding universe, as confirmed by the observations \cite{Riess:1998cb,Perlmutter:1998np,Bennett:2012zja,Ade:2013zuv}.
Recent observational data \cite{Bennett:2012zja,Ade:2013zuv} supports the potential explanation that this acceleration is caused by an exponentially small positive cosmological constant. However, the question remains how this positive vacuum energy is generated.

String theory is a strong candidate to reconcile gravity and its quantum formulation. Therefore string theory may give a hint on how to approach the mystery of the positive cosmological constant, or the realization of de Sitter ($dS$) vacua.
Flux compactifications~\cite{Giddings:2001yu,Dasgupta:1999ss} of 10D string theories are expected to realize a four-dimensional space-time consistent with our universe. After compactification, a choice of quantized fluxes determines the shape of a moduli potential (see reviews \cite{Douglas:2006es,Grana:2005jc,Blumenhagen:2006ci}). Since there exists an exponentially large amount of flux choices, the resultant moduli potential describes the string theory landscape.

It is suggested that the exponential abundance of quantized fluxes allows the realization of a tiny positive cosmological constant \cite{Bousso:2000xa,Feng:2000if}.
However the analysis in \cite{Bousso:2000xa,Feng:2000if} does not take moduli stabilization into account and the resultant distribution of quantized fluxes does not imply the preference of a small cosmological constant.
The dynamics to realize metastable vacua in the moduli potential change the properties of the distribution of the cosmological constant.
Recent studies of the probability distribution of the vacuum energy suggest that a small cosmological constant is preferred in some stringy motivated models in type IIB, owing to the moduli dynamics that correlate terms in the potential to each other \cite{Sumitomo:2012wa,Sumitomo:2012vx,Sumitomo:2012cf,Sumitomo:2013vla}.

Conservatively, vacua with a positive cosmological constant are most likely to arise from extremal points of the moduli potential that satisfy the condition of positivity of all eigenvalues of the mass matrix, i.e., a positive definite mass matrix since they are the most likely to survive uplifting.
In general, the positivity of the mass matrix is an additional strong constraint to extremal points (see e.g., \cite{Shiu:2011zt}), and thus there are many unstable vacua.
Due to the fact that string theory typically has many moduli, $\mathcal{O}(100)$ in the case of Calabi-Yau compactifications, it is an open question if we can satisfy the positivity condition of the mass matrix, especially when we have a large rank of the mass matrix, i.e., a large number of moduli fields.

One may approximate the mass matrix or Hessian
\footnote{The positivity of mass matrix equals to the positivity of Hessian at extremal points. Note that the K\"ahler metric is positive definite for the type of compactifications we are considering in this work: Swiss-Cheese compactifications at large volume. Thus the conversion matrix with respect to field redefinition does not change the positivity of the mass matrix.
}
at the extremal points to obey a Gaussian Orthogonal Ensemble (GOE) owing to the complexity of the stringy setup. Then, the probability that the mass matrix is positive definite was estimated numerically to be a Gaussian suppressed function of the number of moduli \cite{Aazami:2005jf}.
This Gaussian suppression holds even if we change the assumption on the mass matrix towards a more motivated setup in which $dS$ vacua are obtained by uplifting non-tachyonic Anti-de Sitter ($AdS$) vacua \cite{Chen:2011ac}.
The Gaussian suppression on the probability in the case of GOE was further confirmed analytically \cite{Dean:2006wk,Dean2008,Borot:2010tr}.
So when we have a large number of moduli then it may be very difficult to even find one $dS$ vacuum.

In \cite{Ashok:2003gk,Denef:2004ze,Denef:2004cf}, the random Hessian matrix of 4D ${\cal N}=1$ supergravity was proposed, given the superpotential, K\"ahler potential and their derivatives to be random variables
\footnote{See \cite{Giryavets:2003vd,Giryavets:2004zr,Conlon:2004ds,MartinezPedrera:2012rs,He:2013yk}
for counts of the number of vacua in explicit models.
Also a statistical analyses on non-perturbative stability was recently studied in \cite{Greene:2013ida}.}.
Using this setup, the probability for a positive definite Hessian was estimated in detail \cite{Marsh:2011aa,Bachlechner:2012at}.
Especially, at supersymmetric extremal points, the probability is again given as a Gaussian suppressed function of the number of moduli fields \cite{Bachlechner:2012at}.

Supersymmetric extremal points realize $AdS$ vacua which are always stable even if there are tachyonic directions since the mass eigenvalues are given above the Breitenlohner-Freedman bound \cite{Breitenlohner:1982jf,Breitenlohner:1982bm}.
Generically, it is not difficult to achieve stable $dS$ vacua from $AdS$ if the mass matrix is positive definite.
Some of the uplift terms are proportional to a power of total volume, and thus these may not seriously change the structure of the minima of all the moduli as known in many examples.
On the other hand, tachyonic directions in $AdS$ are expected to remain tachyonic in $dS$ for the same reason.

We mentioned above that the probability of stability is Gaussianly suppressed by the number of moduli under the condition that the components of Hessian matrix are given randomly. This is particularly powerful in the case of type IIA supergravity. Here, we generally have to stabilize a large number of moduli simultaneously since we do not have a special configuration which generates a hierarchy for the stabilization scale of different kinds of moduli, unless it is imposed by hand.
Thus it is not easy to achieve stable $dS$ vacua in type IIA as is discussed in \cite{Chen:2011ac}.

On the other hand, in type IIB, we have an approximate `no-scale structure'~\cite{Giddings:2001yu,Ellis:1983sf,Cremmer:1983bf} which splits the K\"ahler sector from the complex structure sector consisting of the complex structure moduli and the dilaton.
This special structure generates a hierarchy so that we can consider the stabilization of each sector separately.
Furthermore, the leading contribution to the potential which dominates the stabilization of the complex structure sector, is convex downward when stabilized supersymmetrically.
Therefore the mass matrix of the complex structure sector is positive definite.
The remaining K\"ahler sector is stabilized by sub-leading contributions to the moduli potential.
In the complete mass matrix, the typical mass scale of the complex structure sector is given higher than that of the K\"ahler sector, which implies a hierarchical structure accordingly.
This hierarchy helps to reduce the effective number of moduli to be considered for stabilization: the complex structure sector with its positive definite mass matrix can be integrated out and only the K\"ahler sector has to be analyzed.
Note that there is an interesting result for a class of models which suggests that the hierarchy is enhanced as the number of complex structure moduli increases \cite{Sumitomo:2012cf}.

In this paper, we study the stabilization of $N_K$ K\"ahler moduli and estimate the probability that the Hessian matrix is positive definite at the extremal points.
We use two classes of models: one is supersymmetric moduli stabilization a la KKLT \cite{Kachru:2003aw}, and the other is the Large Volume Scenario (LVS) type \cite{Balasubramanian:2005zx}.
KKLT introduce a non-perturbative term $W_{\rm NP}$ in the superpotential as a sub-leading contribution to the potential and then stabilize the moduli at supersymmetric extrema.
The hierarchy between the complex structure and K\"ahler sector increases with the overall volume $\Vol$ of the compactification manifold. Hence, in order to have a strong enough hierarchy and also to suppress $\alpha'$-corrections to the moduli potential, the volume is generally demanded to be large. In the KKLT scenario this is achieved by tuning the flux contribution to the superpotential $W_0$ exponentially small.

The LVS \cite{Balasubramanian:2005zx} stabilizes the moduli non-supersymmetrically by combining the leading $\alpha'$-correction of ${\cal O} (\alpha'^3)$ \cite{Becker:2002nn} to the moduli potential with the non-perturbative term $W_{\rm NP}$.
Owing to the absence of a supersymmetric relation, $W_0$ is no longer needed to be tuned small.
In the LVS the volume is related to an exponential of the volume of small four-cycles of a Swiss-Cheese Calabi-Yau compactification such that the volume can become exponentially large easily. 
This helps to avoid contributions from stringy loop-corrections \cite{Berg:2005ja,vonGersdorff:2005bf,Berg:2007wt,Cicoli:2007xp,Cicoli:2008va,Anguelova:2010ed}.
The influence of a recently derived higher order $\alpha'$-correction to the moduli potential~\cite{Grimm:2013gma} was checked  in~\cite{Pedro:2013qga}.

We first systematically analyze the extremal condition in the large volume limit.
As mentioned above, the large volume limit is imposed to suppress $\alpha'$-corrections as well as stringy loop-corrections.
Plugging the extremal condition in the second derivative of the potential, we can obtain a condition for positive definiteness of the Hessian at the extrema in each of the two models respectively.
Interestingly, the condition for positivity of the Hessian turns to be quite simple in both models even though not only the models but also the imposed extremal conditions are quite different: $x_i > 1/4$ in the supersymmetric model, while $x_i > 1$ in the LVS type model.
Here, $x_i = a_i t_i$ for $i=2,..,N_K$ where $t_i$ is the volume of a small four-cycles and $a_i = 2\pi/n_i$ where $n_i$ is the rank of the gauge group that realizes gaugino condensation as a non-perturbative effect on the corresponding cycle.
Next, we scan through a reasonable range of compactification parameters such as the intersection number and flux induced parameters assigned to the each model, and then solve the extremal conditions.
After classifying the solutions by the positivity of the Hessian and imposing a minimal volume constraint to make our supergravity approximations trustable, we can estimate the probabilities for stability.

For a reasonable range of parameters, we observe that the probability of stability for the supersymmetric model is generically higher than the probability for the LVS type model at each value of $N_K$ respectively. This is true if the lower bound for the volume which we impose is the same or even slightly different in both models. 
This tendency is expected to be enhanced as we increase the number of moduli $N_K$ since more unstable solutions would exist in the LVS type model because of the complexity of the extremal conditions in this model compared to the supersymmetric model. Since our analysis is limited by computing time, we restrict to $N_K\leq 6$.

This paper is organized as follows.
In section \ref{sec:supersymm-multi-kahl} we study the supersymmetric model.
We derive the extremal condition for the multi-K\"ahler supersymmetric model and then analyze the positivity condition for the Hessian at the extrema in section \ref{sec:positivity-hessian-kklt}.
Then we give random values for the parameters of the model and estimate the probability of stability in section \ref{sec:numer-simul-kklt}. 
We focus on the LVS type model in section \ref{sec:multi-kahler-lvs}.
Similarly to the supersymmetric model, the extremal and Hessian positivity conditions are obtained in section \ref{sec:positivity-hessian-lvs}.
Then in section \ref{sec:numerical-simulation-lvs}, we calculate the probability and compare our results with the supersymmetric model. Some of the technical details are addressed in the appendixes~\ref{sec:simple-example-ratio} and~\ref{LVSunique}.

\section{Supersymmetric multi-K\"ahler stabilization\label{sec:supersymm-multi-kahl}}
First we consider multi-K\"ahler moduli stabilization by solving supersymmetric extremal equations.
The model is given by
\begin{equation}
 \begin{split}
  K=& -2 \ln {\cal V},\quad {\cal V} = \gamma_1(T_1 + \bar{T}_1)^{3/2} - \sum_{i=2}^{N_K} \gamma_i (T_i + \bar{T}_i)^{3/2},\\
  W=& W_0 + \sum_{I=1}^{N_K} A_I e^{-a_I T_I}.
 \end{split}
 \label{multi-kklt}
\end{equation}
We assume a Swiss-Cheese type of Calabi-Yau compactification for a systematic way to increase the number of moduli fields. A variety of this Swiss-Cheese types can be found in \cite{Denef:2004dm,Cicoli:2008va,Gray:2012jy,Cicoli:2012vw}.
We use indices $I = 1, 2, \cdots, N_K$ and $i = 2, 3, \cdots, N_K$.
Then the supersymmetric critical point equations $D_{T_I} W = 0$, i.e., the KKLT scenario \cite{Kachru:2003aw}, can be simplified together with the set of solutions $\tau_I \equiv \im T_I = 0$ as
 \begin{equation}
  \begin{split}
   B_i \equiv& {A_i \over A_1} = - \frac{\gamma_i}{\gamma_1}\,\frac{a_1}{a_i}\, e^{-a_1t_1 + a_i t_i} \sqrt{\frac{t_i}{t_1}}\, ,\\
   y \equiv& {W_0 \over A_1} = - e^{- a_1 t_1} \left(1 +  {2\over 3} a_1 t_1  - \sum_{i=2}^{N_K}  \frac{\gamma_i}{\gamma_1}\,\frac{a_1}{a_i}\,\sqrt{\frac{t_i}{t_1}} \left(1+ {2\over 3} a_i t_i\right) \right)\,,
  \end{split}
  \label{extremal condition-kklt}
 \end{equation}
Note that the general axionic solutions are given by $\tau_I=m_I \pi/a_I$ with $m_I \in \mathbb{Z}$. Choosing $m_I \neq 0$ in (\ref{extremal condition-kklt}) simply results in a sign change which we can absorb in $A_I$ such that $A_I$ as well as $W_0$ run from negative to positive values.
\footnote{See the effect of warping and a conformal factor in \cite{Dabholkar:2013via}.}

Interestingly, the extremal conditions (\ref{extremal condition-kklt}) can be solved if we expand in 
$x_1^{-1}$ and use the leading order terms. The solution can be written by
\begin{equation}
 \begin{split}
  x_1 \simeq& - {\cal W}_{-1} \left(3 y/2\right), \\
  x_i \simeq& {1\over 2} {\cal W}_0 \left( 2 \frac{\gamma_1^2}{\gamma_i^2}\,\frac{a_i^3}{a_1^3}\,B_i^2\, x_1\, e^{2 x_1} \right)
  = {1\over 2} {\cal W}_0 \left(- {8  \over 9 y^2} \frac{\gamma_1^2}{\gamma_i^2}\,\frac{a_i^3}{a_1^3}\,B_i^2\, {\cal W}_{-1}^3 (3y/2)  \right),
 \end{split}
 \label{solutions-kklt}
\end{equation}
Here we defined $x_I = a_I t_I$ and used the {\it Lambert-W} function ${\cal W} (z)$ which is a solution of the system
\begin{equation}
 z = {\cal W} e^{\cal W}.
\end{equation}
The subscript classifies the type of the Lambert-W function: the function satisfying ${\cal W} \leq -1$ is defined as ${\cal W} = {\cal W}_{-1}$, while ${\cal W} = {\cal W}_{0}$ for ${\cal W} > -1$.
Since $x_1$ should be positive and $t_1 \gg 1$ for the large volume approximation of our interest, the solution of $x_1$ is given by ${\cal W}_{-1}$.
The solution of $x_i$ $(i\geq 2)$ should be positive as well, thus given by ${\cal W}_0$.

The formula of the solutions (\ref{solutions-kklt}) restricts the range of parameters.
Since the volume should be positive $x_1 > 0$, the fundamental domain of the solution restricts the combined parameter $y = W_0/A_1$ to be in the range
\begin{equation}
 -{2\over 3 e} \leq y < 0\,.
  \label{pararange-y}
\end{equation}
On the other hand, from the definition of the volume in (\ref{multi-kklt}) to be positive, we have to satisfy at least $x_1 > x_i$.
Using the property of the Lambert-W function, the second equation of (\ref{solutions-kklt}) can be rewritten as
\begin{equation}
 {\gamma_1^2 a_i^3 \over \gamma_i^2 a_1^3} B_i^2 = {x_i \over x_1} e^{-2(x_1 - x_i)} < 1,
  \label{pararange-Bi}
\end{equation}
where the last inequality comes from $x_1 > x_i$ in case of $a_I = a_J$ which will be used afterwards.
Furthermore, the first extremal condition in (\ref{extremal condition-kklt}) constrains $B_i$ to be negative.

\subsection{Positivity of Hessian\label{sec:positivity-hessian-kklt}}

Since all solutions here are supersymmetric, the stability in $AdS$ is guaranteed as the tachyonic mass eigenvalues are all above the Breitenlohner-Freedman bound \cite{Breitenlohner:1982jf,Breitenlohner:1982bm}.
However, our interest is the realization of $dS$ vacua after eventually introducing an uplifting term. 
Therefore the tachyonic directions would be problematic by the time we uplift the $AdS$ vacua to $dS$.
We want to answer the following question: what is the probability that all eigenvalues of the Hessian matrix $\partial^2 V$ at extremal points are positive. This is similar to the analyses considered in \cite{Bachlechner:2012at,Sumitomo:2012cf}. In this section, we want to estimate the probability to achieve stable $dS$ vacua if we start from the supersymmetric construction of the model defined in (\ref{multi-kklt}). Our analysis is performed under the assumption that the uplifting term does not significantly change the Hessian.
There are many ways to uplift and it is not certain which uplift can be applied for a given $AdS$ solution. Hence, we simply estimate the probability of a positive definite Hessian matrix in $AdS$. A stability analysis together with a rigid uplift is beyond the scope of this paper and is left for a future work.

Next we calculate the Hessian matrix in the large volume approximation $x_1 \gg 1$.
Plugging the extremal conditions (\ref{extremal condition-kklt}) into the defining formula of the Hessian $\partial_{t_i} \partial_{t_j} V$ of the real directions of the $T_i$ and expanding under the assumption of large $x_1$ (or $t_1$), the leading order terms of each component become
\begin{equation}
 \begin{split}
  \left( A_1^{-2} e^{2 x_1} \right) \partial_{t_1}^2 V|_{\rm ext} \sim& {a_1^5 \over 3 \gamma_1^2 x_1} + \cdots,\\
  \left( A_1^{-2} e^{2 x_1} \right) \partial_{t_i}^2 V|_{\rm ext} \sim& {a_1^{9/2} a_i^{1/2} \gamma_i \over12 \gamma_1^3\, x_1^{5/2} x_i^{1/2}} (2 x_i +1) (4x_i -1) + \cdots,\\
  \left( A_1^{-2} e^{2 x_1} \right) \partial_{t_1} \partial_{t_i} V|_{\rm ext} \sim& -{a_1^{11/2} \gamma_i\, x_i^{1/2} \over2 a_i^{1/2} \gamma_1^3\, x_1^{5/2}} (2x_i+1) + \cdots,\\
  \left( A_1^{-2} e^{2 x_1} \right) \partial_{t_i} \partial_{t_j} V|_{\rm ext} \sim& {a_1^6 \gamma_i \gamma_j\, \over 4 (a_i a_j)^{1/2}} {(x_i x_j)^{1/2} \over x_1^4} (2x_i + 1)(2x_j + 1) + \cdots, 
 \end{split}
 \label{hessian-kklt}
\end{equation}
where $i,j$ run for $i,j \geq 2$ and $i\neq j$.
As is clear from the formula above, the diagonal components determine the positivity of all eigenvalues of the Hessian matrix, while the off-diagonal components appear at sub-leading order.
One can check this by using {\it Sylvester's criterion}; the positivity of the sub-matrices are necessary conditions for the positivity of the entire matrix.
Thus the conditions for positivity of the Hessian matrix read
\begin{equation}
 x_1 \gg x_i \geq {1\over4}
  \label{positivity condition-kklt}
\end{equation}
for $i\geq 2$.

This criterion can be made more precise using \eqref{solutions-kklt}. An unstable solution satisfies
\begin{align}
\begin{aligned}
 &a_i t_i < \frac14\,,\\
 \Leftrightarrow \qquad\quad& {\cal W}_0 \left( 2 \frac{\gamma_1^2}{\gamma_i^2}\,\frac{a_i^3}{a_1^3}\,B_i^2\, a_1 t_1\, e^{2 a_1 t_1} \right) < \frac12\,,\\
 \Leftrightarrow \qquad\quad&  \frac{\gamma_1^2}{\gamma_i^2}\,\frac{a_i^3}{a_1^3}\,B_i^2\, a_1 t_1\, e^{2 a_1 t_1}  < \frac14 e^{1/2}\,,\\
 \Leftrightarrow \qquad\quad& \frac{a_i^3 B_i^2}{\gamma_i^2 y^2} < \frac{9\, e^{1/2}}{2 {\cal V}} \simeq \frac{7.42}{{\cal V}}\,,
\end{aligned}
\end{align}
where in the last step we have used ${\cal V}\simeq \gamma_1 (2 t_1)^{3/2}$. Typically $y$ is exponentially small in order to reach a volume ${\cal V}$ where one can trust the supergravity approximation. Furthermore, there is an additional suppression by $1/{\cal V}$ on the RHS which leads us to the conclusion that unstable vacua are expected to be rare, only occurring for very small $B_i$ and/or $a_i$.

The imaginary sector $\partial_{\tau_i} \partial_{\tau_j} V$ of the Hessian can be estimated similarly.
At large $x_1$, the components can be expanded, after plugging in (\ref{extremal condition-kklt}) together with $\tau_i = 0$, by
\begin{equation}
 \begin{split}
  \left( A_1^{-2} e^{2 x_1} \right) \partial_{\tau_1}^2 V|_{\rm ext} \sim& {a_1^5 \over 3 \gamma_1^2\, x_1} + \cdots,\\
  \left( A_1^{-2} e^{2 x_1} \right) \partial_{\tau_i}^2 V|_{\rm ext} \sim& {a_1^{9/2} a_i^{1/2} \gamma_i x_i^{1/2} \over 6 \gamma_1^3\, x_1^{5/2}} \left(4x_i +3 \right) + \cdots,\\
  \left( A_1^{-2} e^{2 x_1} \right) \partial_{\tau_1} \partial_{\tau_i} V|_{\rm ext} \sim& -{a_1^{11/2} \gamma_i x_i^{1/2} \over 2 a_i^{1/2} \gamma_1^3\, x_1^{5/2}}  (2x_i+1) + \cdots,\\
  \left( A_1^{-2} e^{2 x_1} \right) \partial_{\tau_i} \partial_{\tau_j} V|_{\rm ext} \sim& {a_1^{6}\gamma_i \gamma_j \over 2 (a_i a_j)^{1/2} \gamma_1^4} {(x_i x_j)^{1/2} \over x_1^4} \left(2 x_i x_j + x_i + x_j \right) + \cdots, 
 \end{split}
 \label{hessian-imaginary-kklt}
\end{equation}
where $i,j\geq 2$ and $i\neq j$.
Hence, the positivity of the imaginary sector of the Hessian is guaranteed as long as $x_1 \gg 1$ and $x_1 \gg x_i$.

\subsection{Numerical simulation \label{sec:numer-simul-kklt}}
We introduce random parameters to estimate the probability that all eigenvalues of the Hessian matrix become positive.
In a generic Calabi-Yau compactification, the parameters $y=W_0/A_1, B_i=A_i/A_1$, as defined in (\ref{extremal condition-kklt}), are given as complicated functions of the complex structure moduli and the dilaton.
In type IIB string theory on Calabi-Yau manifolds with 3-form flux, the complex structure moduli and the dilaton are stabilized at higher energy scales than the K\"ahler moduli. Integrating out the complex structure sector, the stabilized values of these moduli fields are used as an input for the stabilization of the K\"ahler moduli. There is a large variety in complex structure moduli stabilization due to an exponentially large amount of flux choices~\cite{Bousso:2000xa,Feng:2000if}. The ranges of the resultant input parameters for K\"ahler moduli stabilization $y, B_i$ have a distribution as discussed in \cite{Sumitomo:2012wa,Sumitomo:2012vx,Sumitomo:2012cf,MartinezPedrera:2012rs,Danielsson:2012by,Sumitomo:2013vla}.
Due to the large variety in the complex structure sector, in this work, we simply assume that the combined parameters $y, B_i$ are given as random values.

In general, the details of the stabilization of the complex structure sector are rather complicated. Hence for simplicity we assume uniformly distributed parameters. It is expected that a non-trivial distribution for the parameters is generated rather than a uniform distribution which is in fact a quite conservative assumption in the following sense:
Even in the torus inspired model studied in \cite{Sumitomo:2012vx}, the distribution of $W_0$ peaks at zero point and it becomes sharper as $h^{2,1}$ increases, even though the model is quite simple and such a peaking behaviour around zero might not have been expected. Since here we do not specify the model for the complex structure sector, we simply employ conservative uniform distributions.

\subsubsection{Setup\label{sec:setup-kklt}}

Let us discuss the values and ranges for the parameters of K\"ahler moduli stabilization. Since we know that solutions to (\ref{solutions-kklt}) in the large volume limit only exist for $-{2/ 3 e} \leq y \leq 0$ and $-1\leq B_i \leq 0$, this sets the range for the parameters $y$ and $B_i$. Note that even if we give uniformly distributed random values for $W_0, A_I$, say $-1000 \leq W_0, A_I \leq 1000$,
i.e., all possible values of the parameters are equally likely,
$y, B_i$ obey a uniform distribution in the domain of our interest $-{2/ 3 e} \leq y \leq 0,\, -1\leq B_i \leq 0$, following the general property of ratio distributions (see appendix \ref{sec:simple-example-ratio}).

We only want to study solutions where we can trust our supergravity approximation. Hence, we impose a condition $t_1, t_i> 1$, such that $\alpha'$-corrections to the volume moduli are suppressed. This condition can be rewritten by
\begin{equation}
 1\leq t_I={x_I \over a_I} = {n_I(N_K) \over 2\pi} x_I \leq {n_{I,{\rm max}}(N_K) \over 2\pi} x_I,\label{sugraniconstraint}
\end{equation}
where $I =1, i \ (i\geq 2)$. In the third equality, we assumed that the non-perturbative terms are given by $SU(n_I)$ gaugino condensation on D7-branes, i.e., $a_I = 2\pi/ n_I$. The final inequality is obtained by the constraint on the maximal gauge rank to be consistent with tadpole cancellation and holomorphicity \cite{Collinucci:2008pf,Cicoli:2011qg,Louis:2012nb}.
In \cite{Louis:2012nb}, the maximal gauge rank for Calabi-Yau manifolds which are hypersurfaces in complex projective spaces~\cite{Kreuzer:2000xy} was obtained as a function of $N_K$ as
\begin{equation}
 \begin{split}
  \begin{array}[c]{|c||c|c|c|c|c|c|c|c|c|c|}
  \hline
   N_K & 1& 2& 3& 4& 5& 6& 7& 8& 9& ...\\ \hline
   n_{I,{\rm max}} & 14& 26& 36& 44& 54& 62& 62& 72& 98& ...\\ \hline
  \end{array}
 \end{split}
 \label{max-gauge-rank}
\end{equation}
Note that for e.g., Euclidean D3-branes with $a_I=2\pi$, all solutions that satisfy the supergravity constraint (\ref{sugraniconstraint}) $x_I>2\pi$ automatically satisfy $x_I>1/4$ and hence all such solutions are stable. Stable solutions are least abundant for smallest $a_I$, i.e., largest $n_I$. In the following we set all $a_I$ to the smallest possible value, i.e., $a_I = a_J = 2\pi / n_{I,{\rm max}}(N_K)$. This will lead to the smallest probabilities for a positive definite Hessian such that our results can be interpreted as a lower bound for the probability of stability.

Since we are dealing with rather small values $x_i$ one may worry about the suppression of higher instanton effects $e^{-m'_I a'_I T_I}$ with some positive integers $m'_I$ since these are not included in our Ansatz for the potential in \eqref{multi-kklt}. This is a valid concern.
From our choice of the smallest $a_I$, we have $a'_I > a_I$, so we can at least make sure that contributions from these instantons are smaller as $e^{-m'_I a'_I T_I}\ll e^{-a_I T_I}$.
Since the entire analysis in the presence of all higher instantons is quite complicated, we concentrate on the model with just the largest contributing single non-perturbative terms in this paper.

Among the sets of critical points, we can check the positivity condition of the Hessian (\ref{positivity condition-kklt}).
The condition (\ref{positivity condition-kklt}) does not look very restrictive, but in fact it further restricts the sets of solutions to be minima. As it becomes more difficult to satisfy all the constraints simultaneously in (\ref{positivity condition-kklt}) with growing $N_K$, the probability that the critical points satisfy the stability condition decreases as a function of $N_K$.

Finally, let us discuss the role of the parameters $\gamma_I$ which are given as $\gamma_I={\sqrt{2}}/({3\sqrt{\kappa_I}})$ with $\kappa_I$ being the intersection number of the corresponding four-cycle.
We will estimate the probability for different integers $\kappa=\kappa_I$.
Note that for $\gamma_I=\gamma_J$ being all equal the $x_I$ solutions are not affected by the value 
of $\gamma_I$, see (\ref{solutions-kklt}). However, the value of $\gamma_I$ affects the resultant magnitude of the volume ${\cal V}$ as is clear from (\ref{multi-kklt}).

\subsubsection{Probability \label{sec:probability-kklt}}

We employ a constraint $\Vol > \Vol_{\text{min}}$ in order to make our supergravity approximation trustable.
We start with ${\cal V}_{\rm min}=30$ and increase ${\cal V}_{\rm min}$ gradually to see the effect of this constraint.

If we use the constraint ${\cal V} > 30$ then the probability of stability, i.e., the probability that all eigenvalues of the Hessian turn out to be positive is given by
\begin{equation}
 \begin{split}
  \begin{array}[c]{|c||c|c|c|c|c|}
  \hline
  N_K & 2& 3& 4& 5& 6\\ \hline
  {\cal P} & 0.997& 0.892& 0.668& 0.381& 0.178\\ \hline
  \end{array}
  \qquad {\cal V}\geq 30, \ \gamma_I = {\sqrt{2}\over 3}.
 \end{split}
 \label{kklttable-v30-kappa1}
\end{equation}

Let us demonstrate the effect of changing $\gamma_I$ while keeping $\gamma_I=\gamma_J$ for all $I,J$ for simplicity. If we increase $\gamma_I$ to be $\gamma_I = \sqrt{2}/3\sqrt{5}$, i.e., $\kappa=5$, then the probability is given as
\begin{equation}
 \begin{split}
  \begin{array}[c]{|c||c|c|c|c|c|}
  \hline
  N_K & 2& 3& 4& 5& 6\\ \hline
  {\cal P} & 0.999& 0.908& 0.702& 0.433& 0.226\\ \hline
  \end{array}
  \qquad {\cal V}\geq 30, \ \gamma_I = {\sqrt{2}\over 3\sqrt{5}}.
 \end{split}
 \label{kklttable-v30-kappa5}
\end{equation}
For $\gamma_I = \sqrt{2}/3\sqrt{15}$, i.e., $\kappa=15$ we find
\begin{equation}
 \begin{split}
  \begin{array}[c]{|c||c|c|c|c|c|}
  \hline
  N_K & 2& 3& 4& 5& 6\\ \hline
  {\cal P} & 1.00& 0.914& 0.788& 0.464& 0.389\\ \hline
  \end{array}
  \qquad {\cal V}\geq 30, \ \gamma_I = {\sqrt{2}\over 3 \sqrt{15}}.
 \end{split}
 \label{kklttable-v30-kappa15}
\end{equation}
Hence, the probability increases as $\gamma_I$ decreases. The difference is just a factor, which does not change the order of magnitude within the considered range of $\kappa$.

Next, we change the value of $\Vol_{\text{min}}$.
When we impose ${\cal V} > 50$ with $\gamma_I = \sqrt{2}/3$, we obtain
\begin{equation}
 \begin{split}
  \begin{array}[c]{|c||c|c|c|c|c|}
  \hline
  N_K & 2& 3& 4& 5& 6\\ \hline
  {\cal P} & 0.998& 0.910& 0.704& 0.407& 0.203\\ \hline
  \end{array}
  \qquad {\cal V}\geq 50, \ \gamma_I = {\sqrt{2}\over 3}.
 \end{split}
 \label{kklttable-v50-kappa1}
\end{equation}
Also for ${\cal V} > 100$,
\begin{equation}
 \begin{split}
  \begin{array}[c]{|c||c|c|c|c|c|}
  \hline
  N_K & 2& 3& 4& 5& 6\\ \hline
  {\cal P} & 1.00& 0.913& 0.803& 0.506& 0.299\\ \hline
  \end{array}
  \qquad {\cal V}\geq 100, \ \gamma_I = {\sqrt{2}\over 3}.
 \end{split}
 \label{kklttable-v100-kappa1}
\end{equation}
So the probability increases when we increase the volume $\Vol_{\text{min}}$.

For comparison, we also show the probability for ${\cal V} > 50$ with $\gamma_I = \sqrt{2}/3\sqrt{5}$:
\begin{equation}
 \begin{split}
  \begin{array}[c]{|c||c|c|c|c|c|}
  \hline
  N_K & 2& 3& 4& 5& 6\\ \hline
  {\cal P} & 1.00& 0.901& 0.800& 0.475& 0.318\\ \hline
  \end{array}
  \qquad {\cal V}\geq 50, \ \gamma_I = {\sqrt{2}\over 3\sqrt{5}}.
 \end{split}
 \label{kklttable-v50-kappa5}
\end{equation}
So the probability here is higher than in both (\ref{kklttable-v30-kappa5}) and (\ref{kklttable-v50-kappa1}) at least for $N_K > 3$.
However, we should note that the probabilities do no change so drastically within the range of constraints considered here.

It is difficult to obtain reliable values for the probabilities at larger values of $N_K$ due to the following computational limitation: The uniformly distributed random value of $y$ has to be rather small in order to match the large volume constraint. Hence, most parameter points in our scan will be excluded by this constraint. As we also employ the constraint $t_i>1$, it becomes difficult to find enough extremal points for large values of $N_K$ that fulfill the complete set of supergravity constraints.

One may also consider what happens when we randomize $\gamma_I$. In this situation, the values of the probability are given roughly in the middle between the values at each edges, i.e., at highest and lowest $\gamma_I$. Furthermore, the probabilities increase slightly if we enhance the constraint for a hierarchy between $x_1$ and $x_i$, which is important for the approximations we employ for our stability analysis. This can be done by imposing a lower bound on $\gamma_I^{-1} a_I^{3/2} {\cal V} = (2x_1)^{3/2} - \sum (2x_i)^{3/2}$.

Finally, we change the $a_I$ parameters which have been defined via $n_I = n_{I,{\rm max}}(N_K)$ so far. We show the probabilities for ${\cal V}\geq 30$ and $\gamma_I = \sqrt{2}/ 3$ for $a_I=2\pi/ n_I$ with independently
uniformly distributed integers $1\leq n_I \leq n_{I,{\rm max}}(N_K)$:
\begin{equation}
 \begin{split}
  \begin{array}[c]{|c||c|c|c|c|c|}
  \hline
  N_K & 2& 3& 4& 5& 6\\ \hline
   {\cal P} & 1.00& 0.996& 0.988& 0.972& 0.948\\ \hline
  \end{array}
  \qquad {\cal V}\geq 30, \ \gamma_I = {\sqrt{2}\over 3}.
 \end{split}
 \label{kklttable-v30-kappa1_randoma}
\end{equation}
As expected, the probabilities are larger than those given in (\ref{kklttable-v30-kappa1}).

Since the changes in the probabilities of stability in changing the supergravity approximation constraints and the input values for the $\gamma_I$ parameters are rather mild, we conclude that the probability values given above capture essential features of the multi-K\"ahler moduli model of supersymmetric moduli stabilization.
Before proceeding, let us briefly summarize the properties of the probability in the supersymmetric stabilization model:
\begin{itemize}
 \item We find that decreasing $\gamma_I$ increases the abundance of stable critical points. In the spirit of deriving a lower bound for the probability of stability, the smallest stability probability is found for the maximal value for all $\gamma_I= \sqrt{2}/3$, i.e., $\kappa_I = 1$, restricting ourselves to integer intersection values.
       
 \item Since we find that increasing the value of $\Vol_{\text{min}}$ also increases the probability of stability, we chose a rather small value $\Vol_{\text{min}}=30$, again, in the spirit of giving a lower limit on the probability of stability: larger $\Vol_{\text{min}}$ will give larger probabilities.
\end{itemize}

\section{Multi-K\"ahler LVS\label{sec:multi-kahler-lvs}}
Next we consider the multi-K\"ahler moduli model defined by
\begin{equation}
 \begin{split}
  K=& -2 \ln \left({\cal V} + \frac{\xi}{2}\right),\quad {\cal V} = \gamma_1(T_1 + \bar{T}_1)^{3/2} - \sum_{i=2}^{N_K} \gamma_i (T_i + \bar{T}_i)^{3/2},\\
  W=& W_0 + \sum_{i=2}^{N_K} A_i e^{-a_i T_i},
 \end{split}
 \label{multi-lvs}
\end{equation}
with $\xi \propto -\chi\, g_s^{-3/2}$, where $\chi$ is the Euler number of the Calabi-Yau manifold. In the exponentially large volume limit~\cite{Balasubramanian:2005zx}, the leading terms of the induced scalar potential after axion extremization are
\begin{equation}
 V \simeq \frac{2\sqrt{2}}{3 \Vol} \left( \sum_{i=2}^{N_K} \frac{a_i^2 A_i^2 \sqrt{t_i}}{\gamma_i}\, e^{-2 a_i t_i} \right) + \frac{4 W_0}{\Vol^2} \left( \sum_{i=2}^{N_K} (-1)^{m_i} a_i A_i t_i\,e^{- a_i t_i} \right) + \frac{3 W_0^2 \xi}{4 \Vol^3}.
\label{VLVS}
\end{equation}
The axion extrema of the potential lie at $\tau_i = m_i \pi/a_i$ for $m_i\in \mathbb{Z}$. The values of the $m_i$ influence the signs of the terms in the second sum of \eqref{VLVS}.
The approximations we employed to obtain the potential above are a large volume ${\cal V} \gg 1$ and small $\alpha'$-correction $\xi/{\cal V} \ll 1$. Since the cross terms $e^{-x_i - x_j}$ come with ${\cal V}^{-2}$, they are subleading in the large volume approximation and therefore neglected in (\ref{VLVS}).

Introducing the parameters
\begin{equation}
 c_i \equiv {A_i \over W_0}, \quad x_i \equiv a_i t_i\,,
\end{equation}
for $i\geq 2$, the potential (\ref{VLVS}) becomes
\begin{equation}
 \begin{split}
  V \sim& W_0^2 \left( \sum_{i=2}^{N_K} {2\sqrt{2} a_i^{3/2} c_i^2 x_i^{1/2} \over 3 \gamma_i {\cal V}} e^{-2 x_i} + \sum_{i=2}^{N_K} (-1)^{m_i} {4 c_i x_i \over {\cal V}^2} e^{-x_i} + {3 \xi \over 4 {\cal V}^3}  \right).
 \end{split}
 \label{effpot-LVS}
\end{equation}
The extremal condition $\partial_{t_I} V = 0$ can be simplified to be
\begin{equation}
 \begin{split}
  (-1)^{m_i} c_i =& - e^{x_i} {6\sqrt{2} \gamma_i \, x_i^{1/2}(x_i-1) \over a_i^{3/2} {\cal V} (4 x_i-1)},\\
  \xi =&  64 \sqrt{2} \sum_{i=2}^{N_K} { \gamma_i \, x_i^{5/2}(x_i-1) \over a_i^{3/2} (4x_i-1)^2}.
 \end{split}
 \label{extremal-LVS}
\end{equation}
Plugging the condition (\ref{extremal-LVS}) back in the potential (\ref{effpot-LVS}), we obtain
\begin{equation}
 \begin{split}
  V_{\rm ext} =& - W_0^2 \sum_{i=2}^{N_K} {8\sqrt{2} \gamma_i\, (x_i-1) x_i^{3/2} \over  a_i^{3/2} t_1^{9/2} (4x_i-1)^2 }.  
 \end{split}
\end{equation}
So we see that if all $x_i > 1$, then the extremal points always stay in $AdS$.
The extremal solutions may exist even for $x_i < 1$, suggesting the possibility of unstable $dS$ solutions.
But since we are interested in the vacuum stability after uplifting such that all solutions are in $dS$ region, it is fair to compare with all possible extrema for our purpose.

\subsection{Positivity of Hessian\label{sec:positivity-hessian-lvs}}
Now we are ready to discuss the positivity of the Hessian $\partial_{t_I} \partial_{t_J} V$ of the LVS effective potential. Using the extremal conditions (\ref{extremal-LVS}) the Hessian becomes
\begin{equation}
 \begin{split}
  W_0^{-2} \left. \partial_{t_1}^2 V \right|_{\rm ext} \sim& \sum_{i=2} {432 \sqrt{2} \over {\cal V}^{13/3}} {\gamma_i\, x_i^{3/2} (x_i-1) (2x_i+1) \over a_i^{3/2} (4x_i-1)^2}+...\,,\\
  W_0^{-2} \left. \partial_{t_i}^2 V \right|_{\rm ext} \sim& {12 \sqrt{2}  \over {\cal V}^{3}} {a_i^{1/2} \gamma_i\, (x_i-1) (8 x_i^3 - 6 x_i^2 + 3x_i + 1) \over x_i^{1/2} (4x_i-1)^2}+...\,,\\
  W_0^{-2} \left. \partial_{t_1} \partial_{t_i} V \right|_{\rm ext} \sim& - {72\sqrt{2} \over  {\cal V}^{11/3}} {\gamma_i\, x_i (x_i-1)^2 \over a_i^{1/2} x_i^{1/2} (4x_i-1) }+...\,,\\
  W_0^{-2} \left. \partial_{t_i} \partial_{t_j} V \right|_{\rm ext} \sim& {\cal O} \left({\cal V}^{-4} \right)+...\,.
 \end{split}
\end{equation}
The components of $\partial_{t_i} \partial_{t_j} V|_{\rm ext}$ are sub-leading with respect to $\partial_{t_i}^2 V|_{\rm ext}$ and $\partial_{t_1} \partial_{t_i} V|_{\rm ext}$ and thus negligible in the large volume approximation. 
To achieve the positivity of the Hessian matrix, we apply {\it Sylvester's criterion} from the right-bottom corner of the matrix, so $\partial_{t_{N_K}}^2 V|_{\rm ext}$.
The sub-matrix $\partial_{t_i} \partial_{t_j} V|_{\rm ext}$, where $i,j \geq 2$, is diagonal at large volume, therefore the positivity condition requires $x_i > 1$.
Then the remaining condition is obtained just by the positivity of the total determinant.
Since the Hessian has a repetitive structure, the determinant can be simplified to be
\begin{equation}
 \begin{split}
  W_0^{-2 N_K} \det \left. \partial_{t_I} \partial_{t_J} V \right|_{\rm ext} =& {31104 \over {\cal V}^{22/3}} \sum_{i=2} {\gamma_i^2 x_i^2 (x_i-1)^2 \over a_i (4 x_i-1)^4} \left[(12 x_i - 11) x_i + 5 \right] \prod_{j\neq i} W_0^{-2} \partial_{t_j}^2 V|_{\rm ext}.
 \end{split}
\end{equation}
$x_i > 1$ is required for the positivity of $\partial_{t_i}^2 V|_{\rm ext}>0$, and thus the total determinant is shown to be positive as long as
\begin{equation}
 x_i > 1.
\label{stability-LVS}
\end{equation}

Next, we proceed to analyze the remaining part of the Hessian matrix for the imaginary directions of the $T_i$.
The Hessian for the axions $\tau_i$ is given as
\begin{equation}
 \begin{split}
  W_0^{-2} \left. \partial_{\tau_1}^2 V \right|_{\rm ext} \sim & {\cal O} \left(e^{-x_1} \right) +... \sim 0\,,\\
  W_0^{-2} \left. \partial_{\tau_i}^2 V \right|_{\rm ext} \sim & {24 \sqrt{2} \over {\cal V}^{3}} {a_i^{1/2} \gamma_i\, x_i^{3/2} (x_i-1) \over (4x_i-1)}+...\,,\\
  W_0^{-2} \left. \partial_{\tau_1} \partial_{\tau_i} V \right|_{\rm ext} \sim & {\cal O} \left(e^{-x_1} \right) +... \sim 0\,,\\
  W_0^{-2} \left. \partial_{\tau_i} \partial_{\tau_j} V \right|_{\rm ext} \sim & {\cal O} \left({\cal V}^{-4} \right)+...\,.
 \end{split}
\end{equation}
Since we can satisfy $\partial_{\tau_1}^2 V|_{\rm ext} > 0$ easily,\footnote{Note that the potential defined in \eqref{multi-lvs}, implies $\partial_{\tau_1}^2 V|_{\rm ext} = \partial_{\tau_1} \partial_{\tau_i} V|_{\rm ext} = 0$ since we have not included non-perturbative corrections originating from the large volume cycle $T_1$ due to their strong suppression in the large volume limit. However, these corrections give the leading expressions for $\partial_{\tau_1}^2 V|_{\rm ext}$ and $\partial_{\tau_1} \partial_{\tau_i} V|_{\rm ext}$.} the condition for the real sector (\ref{stability-LVS}) also implies positivity of the Hessian in the imaginary directions.
The mixing terms between real and imaginary parts are all zero at the extremal points $\tau_i = m_i \pi/a_i$, therefore the stability analysis for the LVS is completed: An extremal point is stable iff the simple condition (\ref{stability-LVS}) is fulfilled. Note that also in the LVS, one might be worried about higher instanton corrections if $x_i$ is not too large. Hence, the most conservative statement would be that there is always at least the stable $x_i > 1$ solution in the LVS case. This solution is unique and always exists if the supergravity constraints are met, see appendix~\ref{LVSunique}.

\subsection{Numerical simulation\label{sec:numerical-simulation-lvs}}

\subsubsection{Setup\label{sec:setup-lvs}}
We now solve the equations for extrema numerically. While we expect the volume $\Vol$ to be exponentially large it turns out that the blow up cycles are typically given as solutions $x_i \sim \mathcal{O}(1)$. Since it is difficult to solve a system of equations with exponential hierarchies numerically, we first eliminate $\Vol$ from the equations in (\ref{extremal-LVS}).

The first equation in (\ref{extremal-LVS}), $\partial_{t_i} V = 0$ for $i\geq 2$, becomes
\begin{equation}
 x_i^{1/2} = - (-1)^{m_i} {{\cal V} a_i^{3/2} c_i \over 6\sqrt{2} \gamma_i} {4x_i-1 \over x_i-1} e^{-x_i}, 
\end{equation}
whereas $\partial V/ \partial \Vol = 0$ becomes
\begin{equation}
 -\frac{2\sqrt{2}}{3 \Vol^2} \left( \sum_{i=2}^{N_K} \frac{a_i^{3/2} c_i^2 x_i^{1/2}}{\gamma_i}\, e^{-2 x_i} \right) - \frac{8}{\Vol^3} \left( \sum_{i=2}^{N_K} (-1)^{m_i} c_i x_i\, e^{- x_i} \right) - \frac{9 \xi}{4 \Vol^4} = 0. \label{dVdVol}
\end{equation}
Plugging $x_i^{1/2}$ into $x_i^{1/2}, x_i$, but not $e^{-2x_i}, e^{-x_i}$ of \eqref{dVdVol}, we can solve for $\Vol$:
\begin{equation}
{\cal V}^3 = - {27 \xi \over 4} \left( \sum_{i=2}^{N_K} e^{-3 x_i}{a_i^{3} (-1)^{3m_i} c_i^3 x_i (4x_i-1) \over \gamma_i^2 (x_i-1)^2 }\right)^{-1}.
\label{Volofti}
\end{equation}
Finally, we obtain the following equations by plugging (\ref{Volofti}) in (\ref{extremal-LVS}):
\begin{equation}
 \begin{split}
  (-1)^{3m_i} c_i^{3} = e^{3 x_i} {432 \sqrt{2} \gamma_i^3 x_i^{3/2} (x_i-1)^3 \over a_i^{9/2} (4x_i-1)^3} {4 \over 27 \xi} \left( \sum_{k=2}^{N_K} e^{-3 x_k} {a_k^{3} (-1)^{3m_k} c_k^3 x_k (4x_k-1) \over \gamma_i^2 (x_k-1)^2 }\right).
 \end{split}
 \label{numerical-eq-LVS}
\end{equation}
Given the values for parameters, we solve these $N_K-1$ equations for $x_i$, then (\ref{Volofti}) gives the value for ${\cal V}$.

We now estimate the probability of a positive definite Hessian as a function of the parameters $W_0$, $\xi$, $A_i$, $\gamma_i$ and $a_i$. Here again we set $a_i=2 \pi / n_{i,max}$ with $n_{i,max}$ given in (\ref{max-gauge-rank}).
We assume uniform distributions for the parameters $W_0$ and $A_I$, here ranging $-1000 \leq W_0, A_i \leq 1000$. 
Since only $c_i = A_i/W_0$ enters (\ref{numerical-eq-LVS}), the value of the upper/lower bound for $W_0$ and $A_i$ is not crucial for the analysis due to the general properties of ratio distributions discussed in appendix \ref{sec:simple-example-ratio} as long as the distribution is symmetric around zero. Similarly to the supersymmetric model, choosing the minimal value for the $a_i$ corresponds to the minimal probability of stability, i.e., a lower bound on the latter.

As far as the choice of the imaginary solutions is concerned, we can set $m_i=0$ without loss of generality.
Since the $m_i$ always come together with the $c_i$ as the combination in (\ref{extremal-LVS}) and also in (\ref{numerical-eq-LVS}), the distribution of the combined quantities $(-1)^{m_i} c_i$ is the same as in the case where we set $m_i = 0$ for a distribution of the $c_i$ with both positive and negative values equally likely.
Note that odd and even numbers of $m_i$ are on equal footing.\footnote{Alternatively to considering positive and negative $c_i$ we can restrict the $c_i$ to be positive. Then, the solutions with $0 < x_i < 1/4$ or $x_i> 1$ correspond to odd choices of $m_i$ for all $i=2,..,N_K$, as can be seen from (\ref{extremal-LVS}). On the other hand, solutions $1/4 < x_i \leq 1$ correspond to $m_i$ being even. Without loss of generality, we restrict to $m_i = 0$ representing the even and $m_i=1$ representing the odd values. We follow both possible approaches, i.e., $m_i=0$ without loss of generality and $m_i = 0$ or $1$ independently, as a check to our calculations.}

In this model, opposite to the supersymmetric stabilization case, the value of $\gamma_i={\sqrt{2}}/({3\sqrt{\kappa_i}})$ affects the value of the $x_i$ solutions. We choose different integer values of $\kappa=\kappa_i$ such that we can compare with the probability for supersymmetric stabilization under the same conditions.

We also give a random value for $\xi$ since it depends on the string coupling $g_s$ whose value is set by complex structure moduli stabilization. We use the explicit formula of $\xi$:
\begin{equation}
 \xi = - {\zeta(3) \over 4 \sqrt{2} (2\pi)^3} \chi(M) (S + \bar{S})^{3/2}
  \sim 4.85 \times 10^{-3} (N_C - N_K) {g_s}^{-3/2},
  \label{xi-input-lvs}
\end{equation}
where $N_C = h^{2,1}$ is the number of complex structure moduli.
We assign a uniformly distributed random integer value for $1\leq N_C \leq 300$ and a uniform random value for $1 < g_s^{-1} \leq 100$ at each $N_K$.

Now we are ready to solve the equation for the $x_i$ (\ref{numerical-eq-LVS}). Since we do not know what the actual number of solutions to (\ref{numerical-eq-LVS}) is, we have to make sure we numerically find all solutions for a given set of parameters. In the case of $x_i>1$, $\forall\, i\geq 2$ the solution is unique as is shown in Appendix~\ref{LVSunique}. In all other cases, there can exist multiple solutions. As is clear from the first equation of (\ref{extremal-LVS}), if $(-1)^{m_i} c_i < 0$, the solutions may exist in either $ x_i > 1$ and/or $0< x_i < 1/4$, whereas $1/4 < x_i < 1$ for  $(-1)^{m_i} c_i > 0$. Hence, these are the 3 fundamental solution domains. Finding numerical solutions requires the input of a starting point for the $x_i$ for the search algorithm at work. In order to obtain all possible solutions, we separate each of three fundamental solution domains into 6 parts and scan all possible combinations of edges of separated parts as the starting points of numerical solving. We do not see 
significant differences in 
the probabilities even if we employ the separation of 7,8 or 9 parts of each of the solution domains. Thus, a posteriori, we scan a high enough number of starting points to obtain all possible solutions and evaluate the probabilities accordingly.

\subsubsection{Examples and comparison to full potential analysis\label{sec:exampl-comp-full-lvs}}
Before proceeding further with our probability analysis, we illustrate a couple of numerical solutions for a given set of values for the parameters to give the reader a better feeling for our method and approximations.
We consider the following set of parameters at $N_K = 3$:
\begin{equation}
 W_0 = 822, \quad A_2 = -151, \quad A_3 = -766, \quad \xi = 17.5, \quad a_i = {2\pi \over 36}, \quad \gamma_I={\sqrt{2} \over 3}.
\end{equation}
Note that all $c_i=A_i/W_0$ are given negatively.
Then the solutions of (\ref{numerical-eq-LVS}) as well as the corresponding volume (\ref{Volofti}) are
\begin{equation}
 \{{\cal V}, x_2, x_3\}_{\rm approx} \sim \{15.9,\,  2.76 \times 10^{-3},\,  1.27\}, \quad \{12.9,\,  1.05,\,  1.22\}, \quad \{595,\, 0.196,\, 3.37 \}.
  \label{eff-example1}
\end{equation}
Here the first solution does not satisfy the constraint $x_i > a_i \sim 0.175\, (t_i>1)$ and thus is not a physical solution.
For demonstration, we still keep the first solution for the moment
\footnote{The constraint $x_i > a_i$ is an additional constraint to avoid serious stringy correction to the moduli fields measured in the string scale and is not crucial for approximating the effective potential itself.}.
Next, we check whether there exist solutions in the full potential $V_{\rm full} = e^K \left(|DW|^2 - 3 |W|^2\right)$ defined in (\ref{multi-lvs}).
The extremal points of the full potential $\partial_{t_I} V_{\rm full} =0$, which are expected to be modified version of the solutions given in (\ref{eff-example1}), are
\begin{equation}
 \{{\cal V}, x_2, x_3\}_{\rm full} \sim \{64.5,\,  8.71 \times 10^{-2},\, 2.41\}, \quad \{56.7,\,  1.49,\,  2.28\}, \quad \{377,\, 0.190,\, 3.51 \}.
  \label{full-example1}
\end{equation}
The Hessian of the second solution is positive definite since $x_i>1$, as is also suggested by the solutions of approximate potential (\ref{eff-example1}). We see that similar solutions exists even in the full potential and the solutions of the full potential are modified slightly compared to the approximate solutions. The deviation is strongest when the value of $\xi/2$ is not too small compared to the volume.

We also like to show another example, using the set of parameters
\begin{equation}
 W_0 = -878, \quad A_2 = 977, \quad A_3 = -42.1, \quad \xi = 14.9, \quad a_i = {2\pi \over 36}, \quad \gamma_I={\sqrt{2} \over 3}.
\end{equation}
So now we have a positive value for $c_3$ while $c_2$ remains negative.
The solutions for the approximate potential (\ref{VLVS}) are given by
\begin{equation}
 \{{\cal V}, x_2, x_3\}_{\rm approx} \sim \{2200,\,  4.61,\,  0.319\}, \quad \{11.7,\,  1.24,\,  0.989 \}.
  \label{eff-example2}
\end{equation}
The solutions of full potential (\ref{multi-lvs}) given the set of parameters become
\begin{equation}
 \{{\cal V}, x_2, x_3\}_{\rm full} \sim \{2120,\,  4.63,\, 0.321\}, \quad \{54.0,\,  2.24,\,  0.981\}.
  \label{full-example2}
\end{equation}
Again we see agreement between the approximate and full potential.
Here, the Hessian of both solutions has tachyonic directions.

\subsubsection{Probability\label{sec:probability-lvs}}
Solving (\ref{numerical-eq-LVS}) iteratively for a given random set of values for the parameters, we can accumulate a number of solutions at each $N_K$. We take the same volume constraint as we considered in the case of supersymmetric stabilization. On top of that, we exclude all solutions which do not satisfy ${\cal V}>|\xi|/2$ to justify dealing with $\xi$ perturbatively. Also, we only keep solutions that satisfy $x_i > a_i$ so that $t_i > 1$ for the supergravity approximation, similar to the analysis in the supersymmetric stabilization.

When we take the volume constraint to be ${\cal V} > 30$ with $\gamma_i = \sqrt{2}/3$, the resultant probability becomes
\begin{equation}
 \begin{split}
  \begin{array}[c]{|c||c|c|c|c|c|c|}
  \hline
  N_K & 2& 3& 4& 5& 6\\ \hline
  {\cal P} & 1.00& 0.676& 0.230& 0.0332& 0.00458\\ \hline
  \end{array}
  \qquad {\cal V}\geq 30, \ \gamma_i = {\sqrt{2}\over 3}.
 \end{split}
 \label{lvstable-v30-kappa1}
\end{equation}
If we compare with the number obtained in (\ref{kklttable-v30-kappa1}) imposing the same volume constraint, we see that each probability at $N_K\geq 3$ is smaller than in the case of supersymmetric stabilization.
Note that it is clear form the formula (\ref{extremal-LVS}) that for $\xi>0$ all of the extremal points at $N_K=2$ are stable as shown in~\cite{Balasubramanian:2005zx}.

In the following, we show various probabilities for different values of the parameters. We will restrict to $N_k\leq 5$ as the numerical solutions at $N_K\geq6$ require a lot of computation time. First, let us change the value of $\gamma_I$.
For ${\cal V} > 30$ with $\gamma_I = \sqrt{2}/3\sqrt{5}$, i.e., $\kappa=5$, we obtain
\begin{equation}
 \begin{split}
  \begin{array}[c]{|c||c|c|c|c|c|}
  \hline
  N_K & 2& 3& 4& 5\\ \hline
  {\cal P} & 1.00& 0.785& 0.321& 0.0943\\ \hline
  \end{array}
  \qquad {\cal V}\geq 30, \ \gamma_I = {\sqrt{2}\over 3 \sqrt{5}}.
 \end{split}
 \label{lvstable-v30-kappa5}
\end{equation}
When we have a constraint ${\cal V} > 30$ with $\gamma_I = \sqrt{2}/3\sqrt{15}$, the probability becomes
\begin{equation}
 \begin{split}
  \begin{array}[c]{|c||c|c|c|c|c|}
  \hline
  N_K & 2& 3& 4& 5\\ \hline
  {\cal P} & 1.00& 0.854& 0.494& 0.143\\ \hline
  \end{array}
  \qquad {\cal V}\geq 30, \ \gamma_I = {\sqrt{2}\over 3 \sqrt{15}}.
 \end{split}
 \label{lvstable-v30-kappa15}
\end{equation}
As in the case of supersymmetric stabilization, the probability increases gradually as we decrease the value of $\gamma_I$.
However, if we compare with the probabilities of (\ref{kklttable-v30-kappa5}) and (\ref{kklttable-v30-kappa15}) respectively, we see that the probabilities of the LVS always show up with a smaller value.

Next, we check the probability when we increase the volume constraint.
If we take ${\cal V} > 50$ with $\gamma_I = \sqrt{2}/3$, we get
\begin{equation}
 \begin{split}
  \begin{array}[c]{|c||c|c|c|c|c|}
  \hline
  N_K & 2& 3& 4& 5\\ \hline
  {\cal P} & 1.00& 0.633& 0.161& 0.0270\\ \hline
  \end{array}
  \qquad {\cal V}\geq 50, \ \gamma_I = {\sqrt{2}\over 3},
 \end{split}
 \label{lvstable-v50-kappa1}
\end{equation}
and then for ${\cal V} > 100$ with $\gamma_I = \sqrt{2}/3$
\begin{equation}
 \begin{split}
  \begin{array}[c]{|c||c|c|c|c|c|}
  \hline
  N_K & 2& 3& 4& 5\\ \hline
  {\cal P} & 1.00& 0.658& 0.182& 0.0334\\ \hline
  \end{array}
  \qquad {\cal V}\geq 100, \ \gamma_I = {\sqrt{2}\over 3}.
 \end{split}
 \label{lvstable-v100-kappa1}
\end{equation}
Hence, if we increase the lower bound for the volume the probability increases. The increase is rather mild for the following reason: unlike the supersymmetric case, here we have much more larger volume solutions and thus it is not so difficult to impose the larger volume constraint. This simply justifies the common understanding that quite large volume is much easier achieved in the LVS than in the supersymmetric K\"ahler moduli stabilization, given the reasonable range of parameters. Even taking the volume constraint to be some other larger values, e.g., 200, 300, the probabilities stay mostly unchanged.

We also give an example for an intermediate volume constraint and different values of $\gamma_I$.
For ${\cal V} > 50$ with $\gamma_I = \sqrt{2}/3 \sqrt{5}$ we find,
\begin{equation}
 \begin{split}
  \begin{array}[c]{|c||c|c|c|c|c|}
  \hline
  N_K & 2& 3& 4& 5\\ \hline
  {\cal P} & 1.00& 0.793& 0.345& 0.111\\ \hline
  \end{array}
  \qquad {\cal V}\geq 50, \ \gamma_I = {\sqrt{2}\over 3\sqrt{5}}.
 \end{split}
 \label{lvstable-v50-kappa5}
\end{equation}
The probability is still below the one given in (\ref{kklttable-v50-kappa5}).
We also illustrate the case ${\cal V} > 100$ with $\gamma_I = \sqrt{2}/3 \sqrt{15}$:
\begin{equation}
 \begin{split}
  \begin{array}[c]{|c||c|c|c|c|c|}
  \hline
  N_K & 2& 3& 4& 5\\ \hline
  {\cal P} & 1.00& 0.884& 0.623& 0.169\\ \hline
  \end{array}
  \qquad {\cal V}\geq 100, \ \gamma_I = {\sqrt{2}\over 3\sqrt{15}}.
 \end{split}
 \label{lvstable-v100-kappa5}
\end{equation}
Even in this case, the probability is not higher than that of the supersymmetric case under the weakest constraint ${\cal V} > 30$ with $\gamma_I = \sqrt{2}/3$ (\ref{kklttable-v30-kappa1}).

Finally, we consider random values for the $a_i$ by calculating the probabilities for ${\cal V}\geq 30$ and $\gamma_I = \sqrt{2}/ 3$ for $a_i=2\pi/ n_i$ with uniformly distributed integers $1\leq n_i \leq n_{i,{\rm max}}(N_K)$:
\begin{equation}
 \begin{split}
  \begin{array}[c]{|c||c|c|c|c|}
  \hline
  N_K & 2& 3& 4& 5\\ \hline
  {\cal P} & 1.00& 0.908& 0.596& 0.217\\ \hline
  \end{array}
  \qquad {\cal V}\geq 30, \ \gamma_I = {\sqrt{2}\over 3}.
 \end{split}
 \label{lvstable-v30-kappa1_randoma}
\end{equation}
Similar to the case of supersymmetric stabilization, the probabilities are larger than those given in (\ref{lvstable-v30-kappa1}). Again, the probability here is smaller than that in (\ref{kklttable-v30-kappa1_randoma}).

Let us conclude our analysis with some comments:
\begin{itemize}
 \item It turns out that the probability increases once we take into account smaller values $\gamma_I<\sqrt{2}/3$, i.e., larger self-intersections $\kappa_I>1$. Also, as already observed in the case of supersymmetric stabilization in section~\ref{sec:supersymm-multi-kahl}, smaller values for the $a_i$ increase the probabilities. Hence, our result in (\ref{lvstable-v30-kappa1}) presents a lower bound on the stability probability in the LVS for a reasonable range of parameters. Note that even upon increasing the parameter ranges, the LVS probabilities generically stay well below the supersymmetric probabilities. The same effect can be observed when we consider larger values of $\xi$ which corresponds to increasing the upper bound for $g_s^{-1}$ to values greater than 100. We will discuss the range of $\xi$ and its distribution in the next subsection.

 \item One may take into account that $\gamma_I$ is given differently for each of manifold, so it should be randomized as well. When we randomize the value of $\gamma_I$, the probability is given roughly in between those of the highest and lowest $\gamma_I$. So, even though we just present the results for fixed $\gamma_I$, our results essentially capture the relevant features of the system. 

 \item We have to scan all the possible starting points which costs a lot more calculation time than in the case of supersymmetric model. The number of starting points increases exponentially with $N_K$.
 
\end{itemize}

\subsubsection{Comments on different distribution of $\xi$ \label{sec:comm-diff-distr}}

So far we considered as a reasonable input for the $\xi$ parameter (\ref{xi-input-lvs}) a uniformly distributed inverse dilaton with range $1< g_s^{-1} < 100$.
From the second equation of (\ref{extremal-LVS}), it is clear that a positive larger $\xi$ prefers $x_i>1$ solutions more than $x_i<1$ solutions.
Therefore, the range of $\xi$ affects the preference for stable solutions $x_i>1$, and hence the probability.
However, it is also a question if large values of $g_s^{-1}$ as well as $\xi$ are allowed naturally.

To understand this better, we introduce the following setup based on the Gukov-Vafa-Witten superpotential~\cite{Gukov:1999ya}:
\begin{equation}
 \begin{split}
  K_S = - \ln \left(S + \bar{S}\right), \quad   W_0 = C_1 + C_2 S,
 \end{split}
\end{equation}
where $S = g_s^{-1} + i \sigma$ and $C_1$ and $C_2$ depend on the complex structure moduli and fluxes.
Then, as studied in \cite{Rummel:2011cd}, the supersymmetric extremum $D_S W_0 = 0$ is given by
\begin{equation}
 g_s = {C_2 \over C_1}, \quad W_0 = 2 C_1.
\end{equation}
The extremal point for the dilaton is a minimum since its potential is defined convex downward due to the no-scale structure.
When we consider uniform distributions with the range $-500 \leq C_1, C_2 \leq 500$, the minimal values of $W_0$ obey the uniform distribution with $-10^3 \leq W_0 \leq 10^3$.
Then, since now $g_s$ is given as the ratio of uniformly distributed parameters, the distribution of $g_s$ is uniform in its weak coupling domain $0< g_s \leq 1$ as discussed in appendix \ref{sec:simple-example-ratio}.
This implies $1\leq g_s^{-1} < \infty$, but the chance to get larger $g_s^{-1}$ is naturally suppressed by its distribution $P(g_s^{-1}) = 1/(g_s^{-1})^2$.

We now reanalyze the probability using $\xi$ of the form (\ref{xi-input-lvs}), but with this uniformly distributed $0< g_s \leq 1$.
The distribution of the other parameters $W_0, A_i$ remains unchanged and we set the $a_i$ to their minimal value.
The probability is estimated by
\begin{equation}
 \begin{split}
  \begin{array}[c]{|c||c|c|c|c|c|}
  \hline
  N_K & 2& 3& 4& 5\\ \hline
  {\cal P} & 1.00& 0.0677& 0.00978& 0.000569\\ \hline
  \end{array}
  \qquad {\cal V}\geq 30, \ \gamma_i = {\sqrt{2}\over 3}.
 \end{split}
 \label{lvstable-v30-kappa1-gsuni}
\end{equation}
Comparing with the case of the uniformly distributed $1\leq g_s^{-1} \leq 100$ (\ref{lvstable-v30-kappa1}), we see that the probability here is highly suppressed.
This is because the larger values of $\xi$ for uniform $0< g_s \leq 1$ are more unlikely than the previous case, meaning the abundance of unstable solutions is strongly increased.

Even when we employ the previous assumption of uniform $1\leq g_s^{-1} \leq 100$, we already see that the probability of the LVS type is given smaller than that of the supersymmetric model under identical conditions.
Together with the fact that the more motivated distribution of $\xi$ prefers even smaller values of the probability, we conclude that stable vacua in the LVS type model are generically more unlikely than in the supersymmetric model.

\section{Discussions\label{sec:discussions}}

We have discussed the probability that the mass matrix at extremal points is positive definite in both a supersymmetric stabilization model and an LVS type model.
We derived a quite simple condition for positivity of the mass matrix for both of the models analytically.
For a given reasonable range of parameters, we see that the probability in the case of supersymmetric stabilization is given higher than that of the LVS type under identical constraints on the volume.

Although we used a reasonable range of parameters and conceivable distributions, in principle, the parameters and their distributions are determined by the details of the stabilization of the complex structure moduli and dilaton.
So when we combine complex structure moduli stabilization with K\"ahler moduli stabilization using concrete models, we should be able to calculate probabilities that are better motivated from a string theoretical perspective.
However, this analysis seems quite involved, and so we relegate it to future work.

We obtained numerical probability data up to $N_K =5,6$ in the LVS type of model.
The reason we can not go to higher $N_K$ is mainly due to the complexity of numerically solving the extremal conditions.
Even in the supersymmetric model, it is not easy to get data points at larger $N_K$ within our computing resources.
Since the number of $\{N_K,\mathcal{P}\}$ data points is not enough, we have many possible functional forms to fit the data as a function of $N_K$, and thus it is difficult to find a reliable fitting function.
As we explained in the introduction, it would be interesting to see if the probability of the concrete models agrees with a Gaussianly suppressed fitting function, but this is beyond the scope of this paper.

In the two models we have analyzed, all non-tachyonic solutions suggest $AdS$ vacua.
Though we expect that the non-tachyonic directions in $AdS$ remain stable directions when uplifted to $dS$, it remains to be seen whether this is true in concrete examples of uplifting.
There are many ways to uplift the potential from $AdS$ in string theory, for instance explicit SUSY breaking \cite{Kachru:2003aw}, D-term uplifting \cite{Burgess:2003ic} (see \cite{Achucarro:2006zf} for a detailed discussion), and recently proposed dilaton-dependent non-perturbative effects \cite{Cicoli:2012fh} and so on.
It would be interesting to check how the probability changes depending on which scenario of the uplift is employed.

There is a class of models in which $dS$ solutions are realized within a simple framework, known as the K\"ahler Uplift model \cite{Balasubramanian:2004uy,Westphal:2006tn} (see also the systematic formulations in \cite{Rummel:2011cd}).
Although there is a potential concern that the volume in the K\"ahler Uplift model has an upper bound, this upper bound constraint can be relaxed in the Racetrack K\"ahler Uplift model \cite{Sumitomo:2013vla}.
In this paper, we did not analyze these models for the following reason: The comparison of models with a single non-perturbative effect, i.e., the two models studied in this work, and racetrack models, which have more than one non-perturbative effect, may come along with a subtlety. Also the K\"ahler Uplift models suggest $dS$ directly without the necessity of introducing additional uplifting terms.
We hope to report on a fair comparison between the K\"ahler Uplift models and the two models addressed in this paper in the future.

\section*{Acknowledgments}
We have benefited from stimulating discussion with Joseph P. Conlon, Francisco G. Pedro, S.-H. Henry Tye and Alexander Westphal.
YS appreciates the hospitality of the DESY theory group, the organizers of ``String Phenomenology 2013'', and Arnold Sommerfeld Center of LMU Munich, where some parts of this project were done.

\appendix

\section{A simple example of ratio distribution \label{sec:simple-example-ratio}}

Here we illustrate how we can justify $y= W_0/A_1$ and $B_i = A_i/A_1$ obeying uniform distributions as we used in section \ref{sec:numer-simul-kklt}.
We consider the situation that $W_0, A_1, A_i$ are uniformly distributed with the range $-10^3 \leq W_0, A_1, A_i \leq 10^3$, so $P(W_0, A_1, A_i) = 1/2 \times 10^{-3}$.
Then the probability distribution of $y$ can be estimated by the constrained integral:
\begin{equation}
 \begin{split}
  P(y) =& \int_{-10^3}^{10^3} d {W_0} d {A_1}\, P(W_0)\, P(A_1) \, \delta \left( y - {W_0 \over A_1}\right)\\
  =& \left\{\begin{array}{cl}
      {1\over 4} & {\rm for}\  |y|\leq 1,\\
             {1\over 4 |y|^2} & {\rm for}\  1 \leq |y|.
            \end{array} \right. 
 \end{split}
\end{equation}
It is worth commenting that the input parameter range does not affect the resultant formula of $P(y)$ if the range is an interval symmetric around zero.
It is also clear that $P(y)$ is normalized so that $\int_{-\infty}^{\infty} dy\, P(y) = 1$.
Thus, for the range of our interest $-2/3e \leq y \leq 0$ as in (\ref{pararange-y}), the distribution of $y$ is uniformly distributed.
A similar argument holds for $B_i$ for the range of solutions $-1 \leq B_i \leq 0$.
Note that more complicated examples of the ratio distribution as well as product, sum and constrained system are available in \cite{Sumitomo:2012wa}.

\section{Approximate analytical multi-K\"ahler LVS solution} \label{LVSunique}

In the case of $x_i \gg 1$ for all $i\geq 2$, we can give an approximate analytical solution to (\ref{numerical-eq-LVS}). We see from (\ref{extremal-LVS}), that a positive volume $\Vol$ requires the $n_i$ to be odd in this case for positive $W_0$ and $A_i$. Approximating $1-4 x_i \simeq -4 x_i$ and $1- x_i \simeq - x_i$, (\ref{extremal-LVS}) and (\ref{numerical-eq-LVS}) simplify to
\begin{align}
&\frac{\Vol e^{-x_i}}{\sqrt{x_i}} \simeq \frac{3  W_0 \gamma_i}{\sqrt{2} a_i^{3/2} A_i}\,,\label{Vsolsqrttapp}\\
&\sqrt{x_i} \simeq  \frac{a_i^{3/2} A_i \xi^{1/3}}{2^{5/6} \gamma_i} \left(\sum_{j=2}^{N_K} \frac{a_j^3 A_j^3}{\gamma_j^2}\, e^{3(x_i - x_j)} \right)^{-1/3}\,.\label{tiimplicitapp}
\end{align}
We can sort a given set of parameters $\gamma_i$, $a_i$ and $A_i$ by the size of the quantity $\gamma_i / (a_i^{3/2} A_i)$. Since $e^{-x_i} / \sqrt{x_i}$ is monotonically decreasing, the set with the largest $\gamma_i / (a_i^{3/2} A_i)$ will give us the smallest $x_i$ as can be seen from \eqref{Vsolsqrttapp}. Without loss of generality, we assign the label $i=2$ to this case. Now, we can neglect terms that are exponentially suppressed in \eqref{tiimplicitapp} relatively to the leading exponential $e^{3(x_i - x_2)}$ and solve for the $t_i=x_i/a_i$:
\begin{align}
 \begin{aligned}
  \langle t_2 \rangle &\simeq \frac12 \left(\frac{\xi}{2} \right)^{2/3}\,,\\
  \langle t_i \rangle &\simeq \frac{1}{2 a_i} {\cal W}_0\left(\frac{a_i^3 A_i^2 \xi^{2/3} \gamma_2^{4/3}}{2^{2/3} a_2^2 A_2^2 \gamma_i^2}\, e^{2 \langle x_2\rangle}\right)\,.
 \end{aligned}
\end{align}

\bibliographystyle{utphys}
\bibliography{myrefs}

\end{document}